\documentclass[aps,prb,twocolumn,superscriptaddress,floatfix,longbibliography,10pt]{revtex4-2}

\usepackage{amsmath,amssymb,mathtools,bm}
\usepackage{graphicx}
\usepackage{physics}
\usepackage{siunitx}

\usepackage{hyperref}

\begin{document}

\title{Representative-volume sizing in finite cylindrical computed tomography by low-wavenumber spectral convergence}

\author{Fernando Alonso-Marroqu\'in}
\email{fernando@quantumfi.net}
\affiliation{Center for Integrative Petroleum Research, King Fahd University of Petroleum and Minerals, Dhahran 31261, Saudi Arabia}
\affiliation{Department of Computational Physics for Engineering Material, ETH Zurich, 8092 Zurich, Switzerland}

\author{Abdullah Alqubalee}
\affiliation{Center for Integrative Petroleum Research, King Fahd University of Petroleum and Minerals, Dhahran 31261, Saudi Arabia}

\author{Christian Tantardini}
\email{christiantantardini@ymail.com}
\affiliation{Center for Integrative Petroleum Research, King Fahd University of Petroleum and Minerals, Dhahran 31261, Saudi Arabia}

\date{\today}

\begin{abstract}
Choosing a representative element volume (REV) from finite cylindrical Computed Tomography (CT) scans becomes ambiguous when a key field variable exhibits a slow axial trend, which may reflect both genuine geological variability and CT acquisition/reconstruction artifacts, because estimated statistics can change systematically with subvolume size and position rather than converging under simple averaging. A practical workflow is presented for sizing an REV under nonstationary conditions by first suppressing axial drift/trend to obtain a residual field suitable for second-order analysis, and then selecting the smallest analysis diameter for which the low-wavenumber content stabilizes within a prescribed tolerance. The approach is demonstrated on \textit{Thalassinoides}-bearing rocks, whose branching, interconnected burrow networks introduce heterogeneity at length scales comparable to typical laboratory core diameters, making imaging-based microstructural statistics and downstream digital-rock estimates highly sensitive to the chosen subvolume. From segmented data, a scalar ``burrowsity'' field—capturing burrow-related pore spaces and infills—is defined, and axial detrending (with optional normalization) is applied to mitigate acquisition drift and nonstationary trends, while the subsequent covariance/spectral test is evaluated on nested cylinders consistent with the core geometry. Representativeness is then posed as a diameter-convergence problem on nested inscribed cylinders: the two-point covariance and its isotropic spectral counterpart $\widehat{C}$ are estimated, and the smallest diameter at which the low-wavenumber plateau becomes stable is selected. Applied to a segmented \textit{Thalassinoides} core, the method identifies a minimum analysis cylinder of approximately $D_{\mathrm{REV}}\approx 93~\mathrm{mm}$ and $H_{\mathrm{REV}}\approx 83~\mathrm{mm}$, enabling reproducible correlation-scale reporting and connectivity-sensitive property estimation.
\end{abstract}

\maketitle

\section{Introduction}
\label{sec:intro}

Branching three-dimensional (3D) burrow systems assigned to the ichnogenus \textit{Thalassinoides} are widely recognized as characteristic expressions of macroscale bioturbation in shallow-marine sedimentary rocks \cite{Knaust2024, EltomAlqubalee2022, Eltom2021Overlooked}. Its architecture is typically organized as interconnected tunnel networks with frequent T- and Y-junctions, locally forming boxworks or maze-like geometries; depending on depositional conditions and diagenetic overprint \cite{EkdaleBromley2003ComplexThalassinoides,BromleyFrey1974GyrolithesThalassinoides}. These structures may be preserved as open pore spaces, as sediment-filled casts, or as cemented infills\cite{EkdaleBromley2003ComplexThalassinoides,BromleyFrey1974GyrolithesThalassinoides}, collectively referred to here as {\it burrowosity}. In core-scale specimens, a ``mottled'' texture is often reported, in which a comparatively fine-grained host matrix is overprinted by a geometrically complex burrow phase whose characteristic length scales span from the tube diameter
(millimeters) to network connectivity across centimeters
\cite{EkdaleBromley2003ComplexThalassinoides}.

From a petrophysical and geomechanical standpoint, this fabric is not a visual
curiosity. Because the burrow phase is frequently better connected than the
surrounding pore space, preferential flow pathways, directional contrasts, and
strong scale effects in measured transport properties can be induced in
\textit{Thalassinoides}-bearing ichnofabrics. This behavior has been increasingly
documented in reservoir-oriented studies of burrowed media, where a small number
of connected junctions or dominant tunnels can control effective connectivity and
permeability upscaling behavior at the plug-to-core transition
\cite{Eltom2021BurrowGasReservoir, EltomGoldstein2021UpscalingBurrowedReservoirs}.

Direct 3D imaging of such fabrics is enabled by medical computed tomography (CT), and quantitative ``digital rock'' analysis
is thereby facilitated. In contemporary workflows, porosity statistics,
connectivity measures, correlation descriptors, and simulation-ready subdomains
for flow or elastic calculations are routinely extracted
\cite{Andra2013DigitalRockBenchmarks2,Bultreys2016ImagingTransportReview}. A
specific difficulty is, however, encountered in burrowed rocks: the dominant
heterogeneity length associated with the burrow network can be comparable to the
scanned core diameter. In that regime, property estimates are not merely noisy at
small volumes; systematic bias can be introduced by insufficient sampling of
low-wavenumber structure (large-scale connectivity), and by whether a subvolume
includes or excludes a small number of key junctions.

Related developments also emphasize the need to account for sample geometry,
segmentation strategy, and downstream mechanical interpretation. For example,
impact-echo studies on short cylinders illustrate that finite cylindrical
geometry can influence non-destructive characterization protocols
\cite{Pelekis2025ImpactEcho}. Adaptive segmentation methods highlight the
importance of treating locally variable information sequences rather than
assuming a single globally homogeneous model \cite{Lebedev2025AdaptiveSegmentation}.
In rock engineering, digital and numerical stress--strain modeling further
shows that reliable geomechanical predictions depend on representative
geometrical and material descriptions of the rock mass
\cite{Demin2025DigitalModeling}. These works motivate the present emphasis on
a reproducible image-derived support size before correlation-scale or
property-scale quantities are reported.

Accordingly, the selection of a \emph{representative volume} is brought to the
foreground. For a finite cylindrical CT scan, the smallest cylinder
diameter (and corresponding axial extent) is sought such that estimated
statistics of a chosen field (e.g., a burrow/host indicator or a burrowsity
metric) become insensitive, within tolerance, to further increases in sampled
volume. In the porous-media literature, this threshold is commonly referred to as
the \emph{representative elementary volume} (REV) \cite{Bear1972Dynamics}, while
closely related usage in heterogeneous solids and composites often employs the
term \emph{representative volume element} (RVE) \cite{OstojaStarzewski2006MaterialSpatialRandomness,Kanit2003RVE}.
Operationally, representativeness is assessed through finite-size convergence:
low-order descriptors (means and variances, and—once near-stationarity is
enforced—two-point statistics such as covariance or spectrum) are expected to
stabilize as the window size is increased \cite{Torquato2002}. In tomography-based
porous media, this principle is commonly implemented by evaluating nested
subvolumes matched to the acquisition geometry, including cylindrical windows for
core scans \cite{AlRaoushPapadopoulos2010REVmicroCT}. For \textit{Thalassinoides}-bearing samples,
two issues become central: (i) a finite cylinder rather than a periodic box is
imposed by the imaging support, and (ii) pronounced nonstationarity can be
present due to acquisition artifacts and geological trends, which must be
mitigated before covariance- or spectrum-based diagnostics are meaningful
\cite{Schluter2014ImageProcessingMicroCT}.

A useful organization of the problem is obtained by treating REV sizing as a
finite-size convergence question controlled by the low-wavenumber content of the
microstructure: when long-wavelength structure is not adequately sampled, any
estimator of correlation scale (and downstream effective properties) remains
diameter-dependent. This perspective is consistent with the recent
covariance/spectrum-driven sizing analysis of finite computational domains in
stationary random media \cite{TantardiniAlonsoMarroquin2025arXivCapillarity}.
Here, the same logic is adapted to a connectivity-dominated geological fabric
imaged as a segmented CT cylinder, for which (a) a preprocessing route
yielding a near-stationary fluctuation field and (b) a diameter-focused
convergence criterion formulated on nested cylinders are required.

The proposed criterion is complementary to standard REV/RVE approaches.
Classical image-based REV analyses often monitor the convergence of scalar
quantities such as porosity, phase fraction, permeability, elastic moduli,
or variance as the window size increases. Those tests are effective when the
field is approximately stationary and when the target effective property is
known in advance. The present approach instead targets the second-order
structure itself, through the low-wavenumber part of the covariance spectrum.
This is useful for burrowed fabrics because the dominant uncertainty is not
only the mean burrow fraction, but whether the sampled domain contains the
long-wavelength connectivity imposed by tunnel segments and junctions. For
stationary materials without axial trends, the detrending step can be omitted
and the method reduces to a conventional covariance/spectral convergence test
on nested windows.

Concretely, the workflow is composed of two components.

\paragraph{Preprocessing to obtain a near-stationary ``burrowsity'' field.}
Starting from a masked and segmented cylindrical volume, a scalar field is
defined to encode the presence of \textit{Thalassinoides}-related structure (e.g., a
Boolean burrow/host indicator or a derived metric). Slow trends (axial drift, beam
hardening, and other low-frequency components) are then removed and fluctuations
are retained that are approximately weakly stationary over the analysis window,
so that second-order statistics can be interpreted.

\paragraph{A \(\widehat{C}\)-test for selecting an optimal cylinder diameter.}
On nested cylinders of increasing diameter, the two-point covariance (and its
Fourier/Hankel counterpart \(\widehat{C}\)) of the detrended field is estimated
and the stability of low-wavenumber descriptors is monitored. The diameter is
deemed representative once these descriptors converge within tolerance,
indicating that sufficient long-wavelength connectivity information has been
captured and that further enlargement does not materially change inferred
correlation scales.

Overall, a covariance/spectrum-based convergence philosophy is operationalized for
REV sizing in burrowed rocks, while the two practical features that dominate
CT applications—nonstationarity and finite cylindrical support—are
explicitly accounted for.

\begin{figure}
    \centering
    \includegraphics[width=1\linewidth]{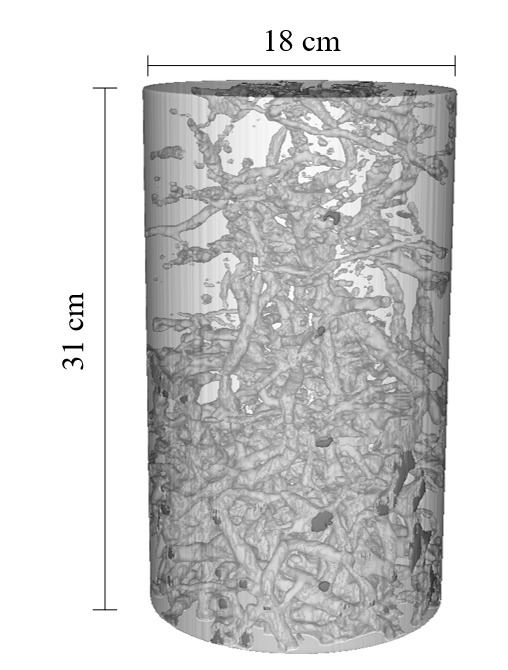}
    \caption{
    Core sample of \textit{Thalassinoides} rock used in the analysis.
    The analyzed CT dataset corresponds to a segmented cylindrical
    core with dimensions \(488\times 488\times 160\) voxels
    (transverse size \(N_x=N_y=488\) pixels and axial length \(N_z=160\) slices).
    The physical specimen dimensions were \(D_{\mathrm{s}}=180~\mathrm{mm}\) in diameter
    and \(H_{\mathrm{s}}=310~\mathrm{mm}\) in height, implying voxel spacings
    \(\Delta x=\Delta y=D_{\mathrm{s}}/488\simeq 0.369~\mathrm{mm/pixel}\) and
    \(\Delta z=H_{\mathrm{s}}/160\simeq 1.94~\mathrm{mm/slice}\).
    }
    \label{fig:placeholder}
\end{figure}

\section{Detrending and denoising of ``burrowsity''}
\label{sec:detrend}

In covariance- or spectrum-based REV criteria, it is implicitly assumed that the
analyzed field is (at least) \emph{weakly stationary} over the window of interest:
the mean is approximately constant and second-order statistics depend primarily on
spatial separation rather than absolute position. In practice, this assumption is
rarely satisfied by raw CT data of geological samples without preprocessing.
Long-wavelength intensity drift, beam hardening, ring artifacts, and genuine
geological gradients can introduce low-frequency components that dominate the
estimated covariance at large lags, thereby inflating inferred correlation lengths
and, consequently, the REV size. Beam-hardening artifacts are a classical source
of such bias in CT, producing cupping and spatially varying attenuation unrelated
to microstructure \cite{BrooksDiChiro1976BeamHardening,BarrettKeat2004ArtifactsCT};
ring artifacts introduce spurious concentric structures that contaminate radial
statistics \cite{SijbersPostnov2004RingArtifacts}. In the present section, an
operational ``burrowsity'' field is defined and a detrending/denoising pipeline is
specified so that a fluctuation field suitable for second-order convergence tests
is obtained.

\subsection{Burrowsity field and detrending to enforce weak stationarity}
\label{subsec:burrowsity_detrend}

The diameter-selection criterion developed in Sec.~\ref{sec:Chat} is formulated
in terms of second-order statistics (covariance and its spectral counterpart),
and a scalar field is therefore required whose fluctuations can be interpreted as
real microstructural variability rather than acquisition drift. In a
\textit{Thalassinoides}-bearing specimen, the most robust choice for this purpose is a
phase-indicator-type field that tracks the burrow-related phase (burrow void,
cast, or cemented infill depending on the segmentation target) on the masked
cylindrical support.

Let $I_{\mathrm{raw}}(\bm{x})$ denote the reconstructed grayscale CT volume,
with voxel-center coordinate $\bm{x}=(x,y,z)$. Let $M(\bm{x})\in\{0,1\}$ be the
cylindrical mask (one inside the core, zero outside), and let the masked intensity
be defined as
\begin{equation}
I(\bm{x}) = M(\bm{x})\,I_{\mathrm{raw}}(\bm{x}), \qquad \bm{x}\in\Omega_{\mathrm{box}},
\end{equation}
so that the effective analysis domain is $\Omega_{\mathrm{cyl}}=\{\bm{x}:M(\bm{x})=1\}$.
In the simplest and most reproducible implementation for REV sizing, \emph{burrowsity}
is defined as the Boolean indicator obtained by thresholding $I$,
\begin{equation}
B(\bm{x})=
\begin{cases}
1, & I(\bm{x})\ge T,\\[2pt]
0, & I(\bm{x})< T,
\end{cases}
\qquad \bm{x}\in\Omega_{\mathrm{cyl}},
\label{eq:booleanB_single}
\end{equation}
where $T$ is a chosen threshold. In practice, $T$ may be selected manually to
match geological interpretation and training slices, or it may be determined by an
objective grayscale criterion such as Otsu’s method for bimodal histograms
\cite{Otsu1979}, with the caveat (well known in porous-media microtomography) that
partial-volume effects, phase overlap in attenuation, and reconstruction artifacts
can render histogram-only criteria insufficient without domain knowledge
\cite{Iassonov2009Segmentation,Schluter2014ImageProcessingMicroCT}.
For REV sizing, Eq.~\eqref{eq:booleanB_single} is intentionally minimal: it is
required primarily that the field be computed \emph{consistently} across nested
cylinders so that observed diameter-dependence can be attributed to sampling and
not to changing preprocessing.

A convenient decomposition is obtained by separating mean phase fraction from
fluctuations. For any analysis domain $\Omega\subseteq\Omega_{\mathrm{cyl}}$,
\begin{equation}
\bar{B}_{\Omega}=\frac{1}{|\Omega|}\sum_{\bm{x}\in\Omega} B(\bm{x}),
\qquad
\delta B(\bm{x})=B(\bm{x})-\bar{B}_{\Omega}.
\label{eq:mean_fluct_single}
\end{equation}
The subsequent steps are designed so that $\delta B$ is not dominated by slow drift
(nonstationary mean) or by structured artifacts that would otherwise leak into the
low-wavenumber part of the covariance/spectrum.

In burrowed rocks, long-wavelength structure may be geological (layering,
cementation gradients, preferential burrow orientation), but in CT data it may
also be produced by acquisition and reconstruction artifacts. This distinction is
critical because the $\widehat{C}$-test is driven precisely by \emph{low} wavenumbers:
any spurious drift or structured artifact injects artificial long-range correlations
and biases the inferred diameter upward.

In the present \textit{Thalassinoides} core sample, the variations of
\(\phi(z)\) reflect genuine geological heterogeneity, including changes in
burrow density, burrow connectivity, infill, or cementation.
The detrending step is therefore used operationally: it separates the large-scale first-order trend
from the residual field used for second-order REV analysis. The removed trend
is not assumed to be unphysical; rather, it represents sample-scale
nonstationarity that would otherwise contaminate the covariance and
low-wavenumber spectrum.

Three effects are particularly relevant.

(i) \emph{Beam hardening and cupping.} In polychromatic CT, preferential attenuation
of low-energy photons produces an effective hardening of the beam with path length.
In reconstructed volumes this commonly appears as a radial intensity gradient
(cupping) and other position-dependent biases
\cite{BrooksDiChiro1976BeamHardening,BarrettKeat2004ArtifactsCT}. For natural
materials with strong heterogeneity, dedicated beam-hardening correction procedures
are often required to avoid bias in quantitative descriptors
\cite{KetchamHanna2014BeamHardeningCorrection}.

(ii) \emph{Ring artifacts.} Detector nonuniformities and reconstruction imperfections
can generate concentric rings in slices. Because the statistics used here are
radial/cylindrical by design, ring artifacts are especially pernicious: an artificial
radial signature is introduced that can be misread as genuine long correlation length
\cite{SijbersPostnov2004RingArtifacts}. Ring-artifact reduction is therefore not merely
cosmetic; it is a prerequisite when the analysis itself is radial.

(iii) \emph{Slice-to-slice drift and other low-frequency trends.} Scanner instabilities,
reconstruction regularization, and/or genuine geological trends can generate slow axial
drift in the mean intensity and, after segmentation, in the segmented phase fraction.
This issue is widely discussed in pore-scale imaging workflows because it couples
field-of-view, grayscale variability, and apparent representativeness
\cite{WildenschildSheppard2013XrayReview,Singh2020GRL}. If uncorrected, the empirical
covariance of a finite cylinder can exhibit slow decay or apparent plateaus that do not
reflect microstructure.

Other forms of nonstationarity may also occur in CT data. These include
local threshold drift caused by grayscale overlap between phases, partial-volume
effects near phase boundaries, mineralogical or cementation gradients, bedding,
fracture-controlled fabric changes, moisture- or contrast-related intensity
variations, and reconstruction artifacts associated with off-axis positioning
or imperfect masking. The present detrending strategy is designed mainly for
slow first-order trends and, when Eq.~\eqref{eq:standardize_single} is used,
for slow variance drift. Abrupt facies changes or sharp stratigraphic
transitions should not be homogenized by a single detrending operation; in such
cases, the analysis should be performed separately within each facies or
structural domain.

To isolate microstructural fluctuations from slow axial drift, the analysis is begun with
the slice-wise phase-fraction signal
\begin{equation}
\begin{aligned}
\phi(z_k)
&=\big\langle B(x,y,z_k)\big\rangle_{x,y} \\
&=\frac{1}{|\Omega_{xy}(z_k)|}
\sum_{(x,y)\in\Omega_{xy}(z_k)} B(x,y,z_k)\,,
\end{aligned}
\label{eq:phi_z_single}
\end{equation}

where $\Omega_{xy}(z_k)=\{(x,y):M(x,y,z_k)=1\}$ is the in-slice support. A smooth trend
$\mu_w(z_k)$ is then constructed by a moving-average (rolling-mean) filter of width $w$
slices,
\begin{equation}
\mu_w(z_k)=\frac{1}{w}\sum_{j=k-\lfloor w/2\rfloor}^{k+\lfloor w/2\rfloor}\phi(z_j),
\label{eq:rolling_single}
\end{equation}
with endpoint handling implemented by truncation or reflection in practice. The detrended
field is defined by subtracting the axial trend \emph{slice-wise},
\begin{equation}
b(\bm{x}) = B(\bm{x})-\mu_w(z), \qquad \bm{x}\in\Omega_{\mathrm{cyl}},
\label{eq:b_field_single}
\end{equation}
so that (approximately) zero mean across $z$ is enforced while the burrow-network geometry
in $(x,y)$ is preserved.

Other detrending operators could be used in the same framework. Polynomial
or spline detrending, LOESS smoothing, Savitzky--Golay filtering, Fourier
high-pass filtering, or morphology-based background correction would all
produce a residual field suitable for the subsequent covariance test, provided
that the imposed cutoff scale is larger than the dominant burrow-scale
intermittency. We used a moving-average trend because it is controlled by a
single interpretable length \(w\Delta z\), is straightforward to reproduce,
and does not impose a global polynomial shape on the axial profile. The
important point for REV sizing is not the particular filter itself, but the
separation between slow nonstationary drift and residual microstructural
fluctuations. In practice, the selected \(D_{\mathrm{REV}}\) and
\(H_{\mathrm{REV}}\) should be reported together with the detrending window
and, when possible, with a sensitivity check using nearby values of \(w\) or
an alternative smooth trend estimator.

When the variance is also observed to drift with $z$ (e.g., due to changing contrast or
layering), an additional slice-wise standardization can be applied,
\begin{align}
\tilde b(\bm{x})&=\frac{B(\bm{x})-\mu_w(z)}{\sigma_w(z)+\varepsilon}, \\
\sigma_w^2(z_k)&=\Big\langle \big(B-\mu_w(z_k)\big)^2\Big\rangle_{x,y},
\label{eq:standardize_single}
\end{align}
where $\varepsilon$ prevents numerical issues in near-constant slices. This normalization
is intended to target \emph{weak stationarity} over the analysis window: a nearly constant
mean and a covariance that depends primarily on separation, not on absolute position. Since
the downstream criteria are second-order, this is the relevant notion of stationarity for
the $\widehat{C}$-based diameter test \cite{AriasCalluari2022Testing}.

The normalization in Eq.~\eqref{eq:standardize_single} is optional and should
be used only when the residual variance varies systematically along the core,
for example because of contrast drift, changing segmentation confidence, or
strong layering. Its primary effect is to make the subsequent covariance
comparison sensitive mainly to correlation shape and length scale rather than
to slice-dependent amplitude. However, if the variance variation is itself a
physically meaningful part of the fabric, normalization may remove information
that is relevant for property prediction. For this reason, the unnormalized
detrended field \(b(\mathbf{x})\) is used as the default, while the normalized
field \(\tilde b(\mathbf{x})\) is recommended as a sensitivity test when
variance drift is evident.

Finally, the choice of the window size $w$ is required to be large enough to remove slow drift but not so
large that genuine burrow-scale intermittency is suppressed. A compact diagnostic is provided
by the excess kurtosis of the detrended axial signal $r(z_k)=\phi(z_k)-\mu_w(z_k)$,
\begin{align}
K_{\mathrm{ex}}(w)&=\frac{m_4}{m_2^2}-3, \label{eq:kurtosis_single} \\
m_p&=\frac{1}{M}\sum_{k=1}^{M}\big(r(z_k)-\bar r\big)^p,
\end{align}
which tracks deviations from Gaussian-like residual fluctuations as $w$ is varied
\cite{AriasCalluari2022Testing}. In Sec.~\ref{sec:detrend}, this criterion is used,
together with covariance/spectrum consistency checks, to select a window width that yields
a residual field suitable for the subsequent $\widehat{C}$-test on nested cylinders.

\subsection{Detrending strategy: axial trend removal and weak-stationary residual}
\label{subsec:detrend_strategy}

Having defined the burrowsity field and identified the dominant sources of
nonstationarity (beam hardening/cupping, ring artifacts, and slow axial drift),
the detrending operator used to produce an analysis field suitable for second-order
(covariance/spectral) convergence tests is specified in this subsection.

Because the acquisition domain is a finite cylinder aligned with the $z$ axis, a
natural and conservative first step is the removal of trends that manifest as slow
variations of the slice-wise phase fraction along $z$. In what follows, it is assumed
that the axial profile $\phi(z_k)$ and the moving-average trend $\mu_w(z_k)$ have
already been defined in Eqs.~\eqref{eq:phi_z_single} and \eqref{eq:rolling_single},
and that the voxel-level detrended field $b(\bm{x})$ has been defined in
Eq.~\eqref{eq:b_field_single}.

The centered moving-average trend in Eq.~\eqref{eq:rolling_single} is applied \emph{only}
to the slice-wise axial profile $\phi(z_k)$, i.e., to the one-dimensional discrete
sequence $\phi_k\equiv \phi(z_k)$ indexed by the slice number $k$ (with spacing $\Delta z$).
Accordingly, the associated transfer function is a \emph{one-dimensional axial}
frequency response. Denoting by $\omega_z$ the discrete axial angular frequency
(in rad/slice; equivalently $\kappa_z=\omega_z/\Delta z$ in rad/mm), the moving-average
kernel of width $w$ has transfer function
\begin{equation}
H_w(\omega_z)
= \frac{1}{w}\,\frac{\sin\!\left(\tfrac{w\omega_z}{2}\right)}{\sin\!\left(\tfrac{\omega_z}{2}\right)}
\,e^{-i\omega_z (w-1)/2},
\label{eq:Hw}
\end{equation}
so that higher-$\omega_z$ content in the \emph{axial} profile is systematically suppressed
as $w$ is increased, while the lowest-$\omega_z$ drift is retained in the trend estimate.
This interpretation is useful for REV sizing because the subsequent $\widehat{C}$-test is
explicitly sensitive to low-wavenumber structure: any residual drift leaking into $b$
inflates apparent correlation scales and biases diameter selection toward overly large
values.

In the present workflow, the primary tuning diagnostic is the residual excess kurtosis
$K_{\mathrm{ex}}(w)$ defined in Eq.~\eqref{eq:kurtosis_single}. To complement this
distributional diagnostic with a second-order criterion aligned with the later covariance
analysis, the normalized autocovariance of the axial residual may also be monitored,
\begin{align}
\widetilde{R}_r(\ell;w)
=&\frac{1}{(M-\ell)\,\sigma_r^2(w)} \nonumber \\
&\sum_{k=1}^{M-\ell}
\big(r(z_k;w)-\overline{r}(w)\big)\big(r(z_{k+\ell};w)-\overline{r}(w)\big),
\label{eq:rr_autocov}
\end{align}
where $\sigma_r^2(w)$ is the sample variance of $r(\cdot;w)$. A compact window-selection
score may then be defined as
\begin{equation}
S(w)=\sum_{\ell=1}^{\ell_{\max}} \left|\widetilde{R}_r(\ell;w)\right|,
\label{eq:stationarity_score}
\end{equation}
with $\ell_{\max}$ chosen as a small fraction of $M$ (e.g., $\ell_{\max}\sim 0.1M$).
In practice, a regime is sought in which both $K_{\mathrm{ex}}(w)$ and $S(w)$ vary weakly
with further increases of $w$, indicating that axial drift has been removed without
artificially erasing intermittent burrow-driven variability.

Axial detrending addresses drift in $\phi(z_k)$ but does not, by itself, remove structured
in-slice artifacts. If rings are present, spurious radial correlation can be introduced
that directly contaminates the radial covariance and spectrum used later. Therefore, when
ring structure is evident by inspection or by strong narrowband peaks in the 2D Fourier
domain of slices, ring-reduction should be performed prior to computing $C$ and
$\widehat{C}$ \cite{SijbersPostnov2004RingArtifacts,BarrettKeat2004ArtifactsCT}. Similarly,
strong beam-hardening/cupping gradients should be corrected during reconstruction or
normalized post hoc to avoid biasing large-scale statistics
\cite{BrooksDiChiro1976BeamHardening,KetchamHanna2014BeamHardeningCorrection}.

All subsequent diameter tests are formulated in terms of second-order statistics (covariance
and spectral descriptors). Accordingly, the operational objective of the present preprocessing
is that the detrended field $b$ exhibits an approximately constant mean over the analysis
window and that its covariance is governed primarily by spatial separation (weak stationarity
within estimation uncertainty). The $\widehat{C}$-based diameter convergence test in the next
section is computed exclusively on $b$.

\begin{figure}
    \includegraphics[width=\linewidth, trim=1cm 5cm 1cm 5cm, clip]{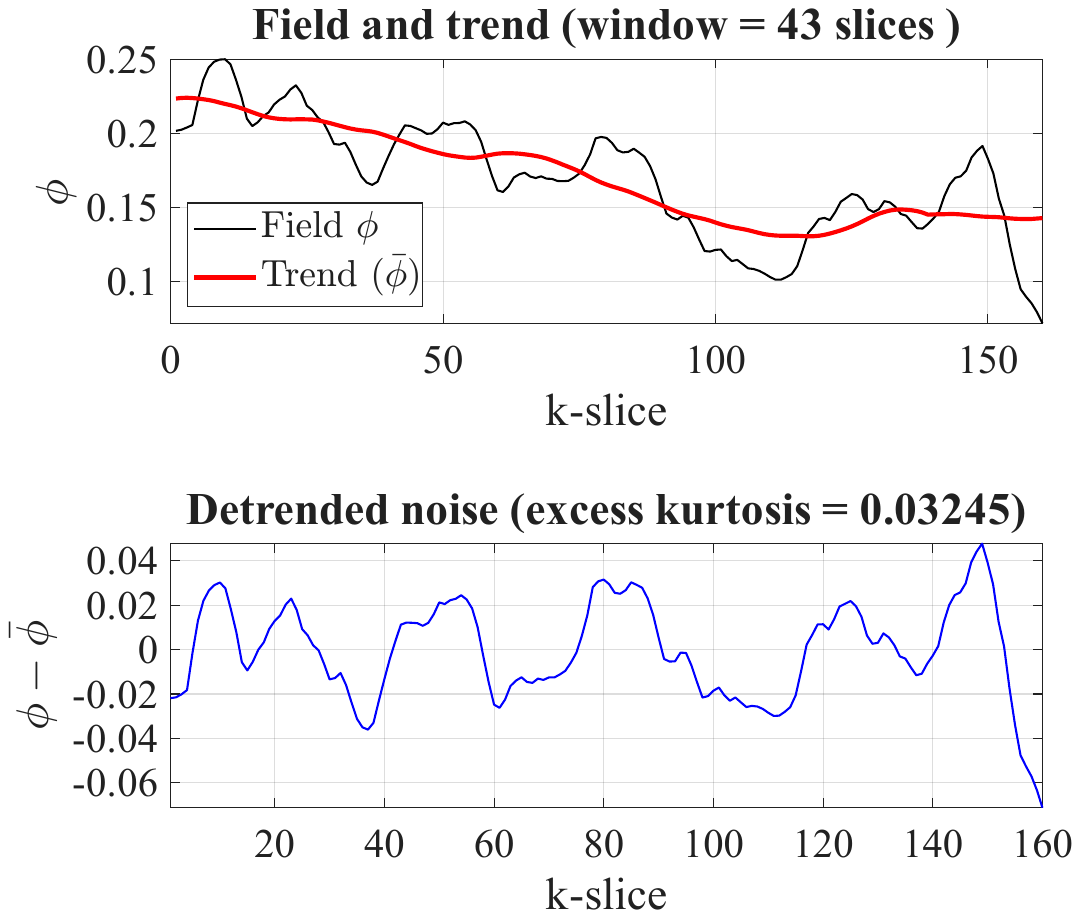}
    \caption{Axial detrending of the slice-wise phase-fraction signal.
    Top: slice-wise phase fraction $\phi(z_k)$ (black) and the corresponding moving-average trend
    $\bar{\phi}(z_k)$ (red) computed with window width $w$ (in slices).
    Bottom: residual $r(z_k)=\phi(z_k)-\bar{\phi}(z_k)$ (blue), used to quantify how much low-frequency
    drift remains after detrending and to guide the choice of $w$ via residual diagnostics (e.g., excess kurtosis).}
    \label{fig:rolling window}
\end{figure}

\begin{figure}
    \includegraphics[width=\linewidth, trim=1cm 4cm 1cm 5cm, clip]{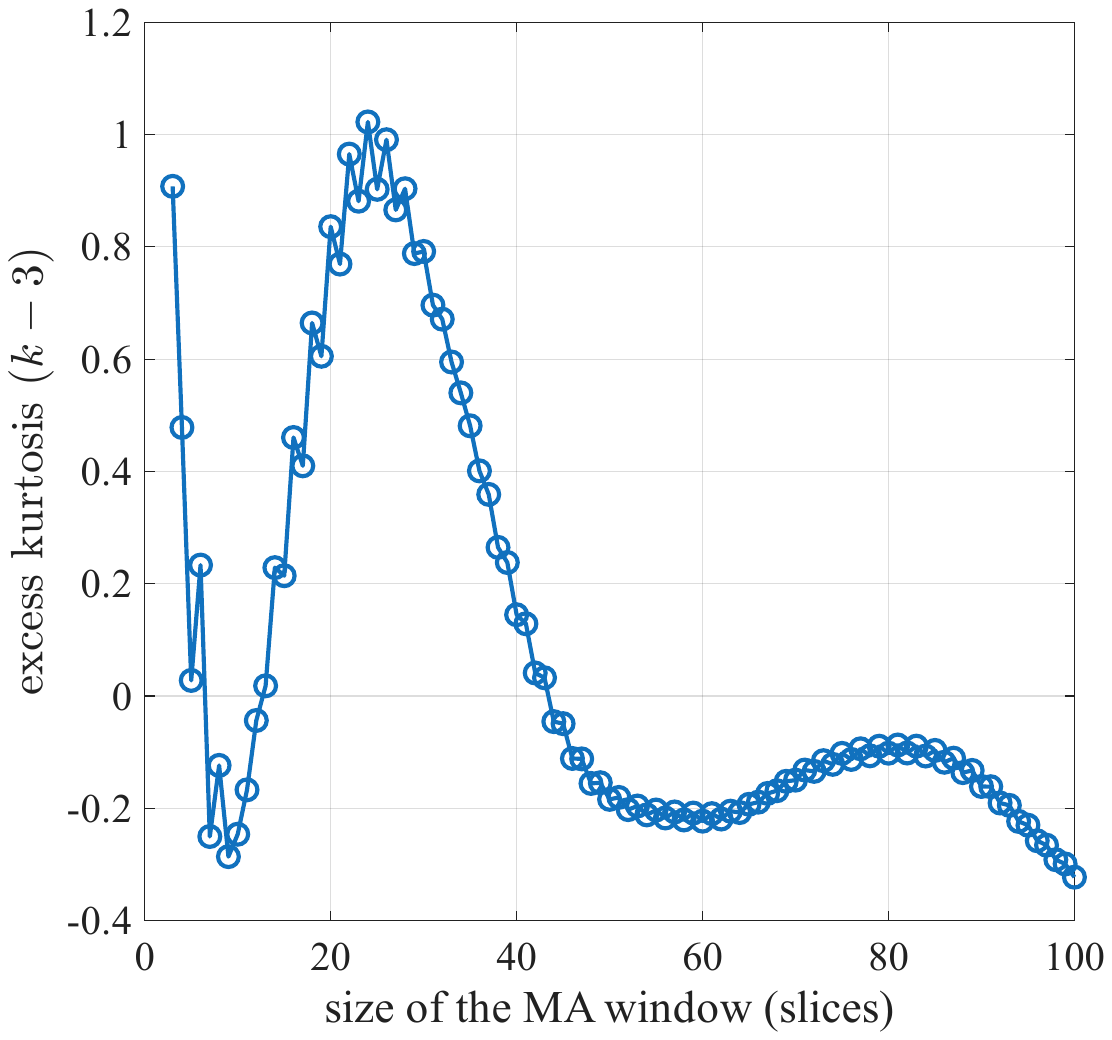}
    \caption{Excess kurtosis $K_{\mathrm{ex}}(w)$ of the detrended axial residual as a function of the moving-average window width $w$.
Because a Gaussian residual has $K_{\mathrm{ex}}=0$, zero-crossings of $K_{\mathrm{ex}}(w)$ are used as a practical criterion for selecting a detrending scale.
Among the candidate windows where $K_{\mathrm{ex}}(w)\approx 0$, the largest window is chosen to suppress slow axial drift most strongly while maintaining a residual that is approximately Gaussian.}
    \label{fig:kurtosis}
\end{figure}

Equations~\eqref{eq:phi_z_single}, \eqref{eq:rolling_single}, \eqref{eq:b_field_single}, and
\eqref{eq:kurtosis_single} define (i) the axial phase-fraction profile, (ii) a smooth trend
obtained by a moving-average window of width $w$, (iii) the voxel-level detrended field, and
(iv) residual diagnostics used to select $w$. Figure~\ref{fig:rolling window} illustrates the
decomposition for a representative choice of $w$, while Fig.~\ref{fig:kurtosis} summarizes the
window sweep used to identify a robust trend/noise separation scale. Through this selection,
the axial extent over which the residual field can be treated as approximately weakly stationary
for the second-order analysis performed below is fixed.

\section{$\widehat{C}$-test for optimal diameter}
\label{sec:Chat}

Through the preprocessing described in Sec.~\ref{sec:detrend}, a detrended residual
field $b(\bm{x})$ [Eq.~\eqref{eq:b_field_single}] is obtained in which slow drift and
structured acquisition artifacts have been reduced to the extent required for
second-order analysis. In the present section, a geometric and statistical objective
is pursued: for a finite CT cylinder, the smallest diameter is determined for
which second-order descriptors of $b$ are stable within a prescribed tolerance.

\subsection{From disk averages to a spectral stability criterion}
As an empirical indicator for representativeness, the stabilization of spatial averages
under increasing window size is often employed. On a cross section at $z=z_k$ (or on a
short axial window centered on $z_k$), the slice-restricted residual is introduced as
\begin{equation}
b_k(\mathbf{x}_\perp)=b(\mathbf{x}_\perp,z_k),
\qquad \mathbf{x}_\perp=(x,y),
\end{equation}
and is defined on the in-slice disk $\|\mathbf{x}_\perp\|\le R$ imposed by the cylindrical
mask. The concentric disk average of $b_k$ over radius $r$ is then given by
\begin{equation}
\langle b_k\rangle(r)=\frac{1}{\pi r^{2}}\int_{\|\mathbf{x}_\perp\|\le r} b_k(\mathbf{x}_\perp)\,d\mathbf{x}_\perp,
\qquad 0<r\le R.
\label{eq:disk_average}
\end{equation}
For small $r$, $\langle b_k\rangle(r)$ is dominated by individual burrow segments and
junctions. As $r$ is increased, disk averaging is effectively applied as a low-pass
filter, and progressively higher spatial frequencies are attenuated.

In the spectral domain, this filtering interpretation is made explicit. Under the working assumption that $b_k$ may be treated as approximately weakly stationary
\emph{within the transverse cross section} at fixed $z=z_k$, the variance of the disk
average can be expressed as a weighted integral of the \emph{isotropic transverse} power
spectrum $\widehat{C}_\perp(k_\perp)$ associated with the two-point covariance of $b_k$,
\begin{equation}
\mathrm{Var}\!\left[\langle b_k\rangle(r)\right]
= \frac{1}{2\pi}\int_{0}^{\infty} \widehat{C}_\perp(k_\perp)\,\big|W_r(k_\perp)\big|^{2}\,k_\perp\,dk_\perp,
\label{eq:var_disk_avg}
\end{equation}
where $k_\perp=\|\boldsymbol{k}_\perp\|$ denotes the radial wavenumber in the $(x,y)$ plane and
$W_r(k_\perp)$ is the transfer function of the normalized circular window used in
Eq.~\eqref{eq:disk_average}. For a normalized radius-$r$ disk in 2D,
\begin{equation}
W_r(k_\perp)=\frac{2J_{1}(k_\perp r)}{k_\perp r},
\label{eq:disk_window}
\end{equation}
with $J_1$ denoting the Bessel function of the first kind. The distinction from the
axial detrending filter should be stressed: $H_w(\omega_z)$ [Eq.~\eqref{eq:Hw}] is a
\emph{one-dimensional} transfer function acting on the axial ($z$-direction) profile
$\phi(z_k)$, whereas $W_r(k_\perp)$ is a \emph{two-dimensional transverse} transfer function
acting on in-plane $(x,y)$ fluctuations within each slice.
Equations~\eqref{eq:var_disk_avg}--\eqref{eq:disk_window}
formalize the practical point that the stabilization of disk-averaged observables is controlled
by the low-$k$ (long-wavelength) content of $\widehat{C}(k)$. On this basis, a diameter criterion
is naturally motivated in which convergence is tested directly on $\widehat{C}(k)$, rather than on
any particular choice of averaged observable.

\subsection{Operational definition of the $\widehat{C}$-test on nested cylinders}
Let $\Omega(D)$ denote the concentric inscribed cylinder of diameter $D$ extracted from the
full scan, with a fixed axis and an axial window selected using $w^\star$ as described in
Sec.~\ref{sec:detrend}. For each $D$, an isotropized in-slice covariance $C_D(r)$ of the residual
fluctuations is computed, and its isotropic spectrum $\widehat{C}_D(k)$ is obtained via the
2D Hankel transform
\begin{equation}
\widehat{C}_D(k)=2\pi\int_{0}^{\infty} C_D(r)\,r\,J_0(kr)\,dr,
\label{eq:hankel}
\end{equation}
where $J_0$ is the Bessel function of the first kind. A diameter is declared representative
once the low-wavenumber part of $\widehat{C}_D(k)$ is found to be insensitive to further increases of $D$.

To make the criterion explicit and reproducible, a convergence metric is evaluated over a
prescribed low-$k$ interval $[0,k_c]$:
\begin{equation}
\varepsilon(D_{m})=
\frac{\left(\int_{0}^{k_c}\big[\widehat{C}_{D_m}(k)-\widehat{C}_{D_{m-1}}(k)\big]^2\,dk\right)^{1/2}}
{\left(\int_{0}^{k_c}\widehat{C}_{D_{m-1}}(k)^2\,dk\right)^{1/2}},
\label{eq:Chat_metric}
\end{equation}
and the smallest $D_m$ is selected such that $\varepsilon(D_m)\le \tau$, where $\tau$ is a user-set
tolerance. The cutoff $k_c$ is chosen to probe wavelengths comparable to, and larger than, the
dominant burrow-network connectivity scale, since that regime is typically responsible for the
strongest diameter dependence in burrowed fabrics.

The computational cost depends on how the covariance is evaluated. A direct
pair-counting estimator scales poorly with the number of voxels and is not
recommended for large CT volumes. In practice, the covariance can be computed
efficiently by FFT-based convolution on each masked slice, followed by radial
binning and Hankel/spectral post-processing. The dominant cost is then of
order \(O(N\log N)\) per slice or per analyzed window, rather than quadratic
in the number of voxels. The isotropic averaging and low-\(k\) integration in
Eq.~\eqref{eq:Chat_metric} are comparatively inexpensive. For very large
datasets, the workflow can be accelerated by processing selected axial windows,
using coarser radial bins, parallelizing over slices and diameters, or using
GPU-based FFT routines. Thus, the method is intended to remain practical for
large micro-CT datasets, provided that the covariance step is implemented with
FFT-based or otherwise accelerated estimators.

\subsection{Slice-wise diagnostics and interpretation}
\begin{figure}
    \centering
    \includegraphics[width=1\linewidth,trim=0cm 4cm 1cm 4cm, clip]{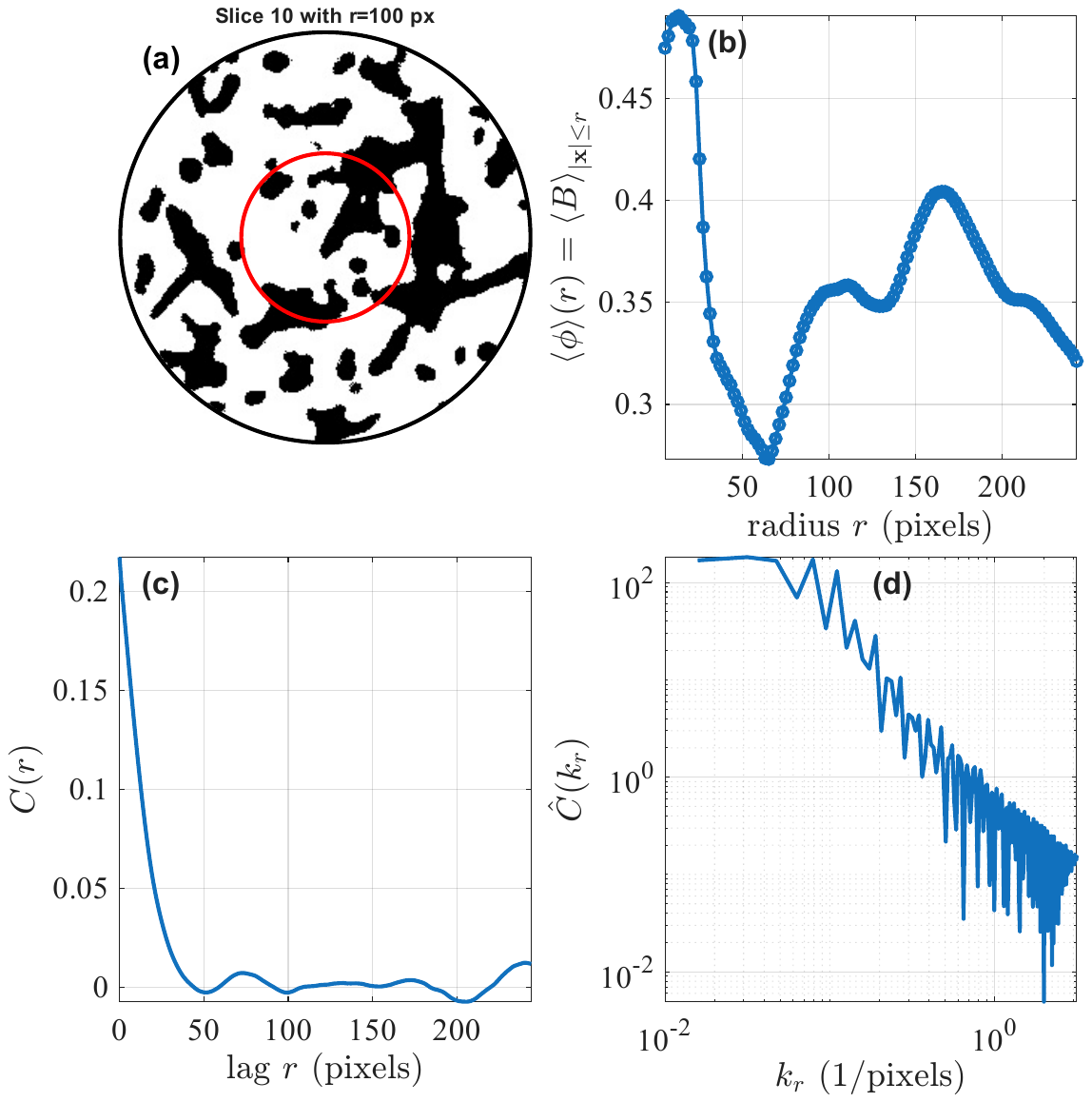}
    \caption{Slice-wise statistics for a \textit{Thalassinoides} core. (a) Boolean slice $B_{\mathrm{slice}}$ (burrow/phase pixels in black, background in white) with a concentric sampling circumference of radius $r$ overlaid in red. (b) Enclosed area fraction $\langle \phi\rangle(r)=\langle B_{\mathrm{slice}}\rangle_{|\mathbf{x}|\le r}$ as a function of radius. (c) Two-point covariance $C(r)$ computed on the slice within an inscribed circular region. (d) Radial spectrum $\widehat{C}(k_r)$ obtained from $C(r)$ using the 2D isotropic Hankel transform.}
    \label{fig:C-hat}
\end{figure}

In Fig.~\ref{fig:C-hat}, complementary readouts are summarized as they are used in practice on a
representative cross section extracted from $\Omega(D)$. In panel (a), a Boolean slice (after masking)
is shown together with the concentric disk used for the average in Eq.~\eqref{eq:disk_average}. In panel
(b), the corresponding enclosed-area fraction (disk-average) is reported as a function of $r$.
In panels (c) and (d), the second-order descriptors used in the $\widehat{C}$-test are provided: the
isotropic covariance $C_D(r)$ and its isotropic spectrum $\widehat{C}_D(k)$ obtained from
Eq.~\eqref{eq:hankel}. In the spectral representation, a transition to a low-$k$ plateau is interpreted
as evidence that the largest resolved wavelengths are being sampled consistently; stability of that
low-$k$ content under increasing $D$ is precisely the condition enforced by Eq.~\eqref{eq:Chat_metric}.
For reporting purposes, a characteristic scale may additionally be extracted from the onset $k_0$ of
the plateau and converted to a radius through $r_{\mathrm{REV}}\sim \alpha/k_0$; the explicit
convergence test above provides the reproducible criterion, while the plateau-based scale serves as an
interpretable diagnostic.

\section{Numerical results}
\label{sec:results}

In this section, the preprocessing protocol described in Sec.~\ref{sec:detrend} and the
$\widehat{C}$-based diameter criterion introduced in Sec.~\ref{sec:Chat} are applied to the
segmented \textit{Thalassinoides} CT cylinder shown in Fig.~\ref{fig:placeholder}. Two objectives
are pursued: (i) dimensionless outputs of the detrending and spectral procedures (window size
in slices, plateau wavenumber in pixels$^{-1}$) are converted into physical REV dimensions, and
(ii) the resulting dimensions are interpreted as practical guidelines for subvolume extraction
and subsequent digital-rock calculations.

\subsection{Voxel spacing and geometric calibration}
The reconstructed volume is defined on a grid of size
$N_x\times N_y\times N_z=488\times 488\times 160$ voxels and corresponds to a physical specimen
of diameter $D_{\mathrm{s}}=180~\mathrm{mm}$ and height $H_{\mathrm{s}}=310~\mathrm{mm}$
(Fig.~\ref{fig:placeholder}). Under the standard linear mapping between pixel counts and specimen
dimensions, the voxel spacings are
\[
\Delta x=\Delta y=\frac{D_{\mathrm{s}}}{N_x}\simeq \frac{180}{488}\approx 0.369~\mathrm{mm/pixel},
\]
\[
\Delta z=\frac{H_{\mathrm{s}}}{N_z}\simeq \frac{310}{160}\approx 1.94~\mathrm{mm/slice}.
\]
Anisotropic resolution is therefore present ($\Delta z\gg \Delta x$), as is typical when full-height
scanning is combined with limited axial sampling. This anisotropy motivates the two-stage sizing logic
developed above: the axial extent is determined by an axial stationarity criterion (Sec.~\ref{sec:detrend}),
whereas the transverse diameter is determined by a radial/spectral criterion evaluated on slices
(Sec.~\ref{sec:Chat}).

\subsection{Axial REV height from detrending window selection}
The axial detrending analysis is performed on the slice-wise phase-fraction signal, and a moving-average
window width $w$ (in slices) is selected such that slow drift is removed while intermittent fluctuations
associated with burrow-network heterogeneity are retained. The window sweep summarized in
Fig.~\ref{fig:kurtosis} yields multiple candidate roots/plateau regimes; following the conservative choice
specified in Sec.~\ref{sec:detrend}, the largest robust window, $w^\star=43$ slices, is selected. The
corresponding axial length scale is

\[
H_{\mathrm{REV}} = w^\star\,\Delta z
\approx 43\times 1.94~\mathrm{mm}
\approx 83.3~\mathrm{mm}.
\]

Operationally, $H_{\mathrm{REV}}$ is interpreted as the minimum axial extent over which the detrended
burrowsity residual can be treated as approximately weakly stationary for the subsequent second-order
analysis. In practice, the $\widehat{C}$-test (and any later property estimation performed on extracted
cylinders) should be restricted to axial windows of length at least $H_{\mathrm{REV}}$. A principled
definition of the axial window used when comparing nested diameters is also provided: diameter dependence
is to be assessed at fixed axial extent so that axial drift effects are not conflated with transverse
finite-size effects.

\subsection{Transverse REV radius from the low-$k$ plateau}
A $\widehat{C}$-based criterion was applied so that the long-wavelength content of the
microstructure sampled within a disk cross section was targeted (Sec.~\ref{sec:Chat}). In the present dataset, a
transition to a low-$k$ plateau was observed in the radial spectrum (Fig.~\ref{fig:C-hat}), and the
onset of this plateau was identified at approximately
\[
k_0 \simeq 0.05~\mathrm{pixel}^{-1}.
\]
This onset was interpreted as the wavenumber associated with the dominant connectivity
scale controlling diameter dependence, and an REV radius in pixel units was estimated as
\[
r_{\mathrm{REV}} \approx \frac{2\pi}{k_0}
\approx \frac{2\pi}{0.05}
\approx 125.7~\mathrm{pixels}.
\]
Conversion to physical units was then carried out using $\Delta x$ (Sec.~\ref{sec:results}), giving
\[
r_{\mathrm{REV}} \approx 125.7\,\Delta x \approx 125.7\times 0.369~\mathrm{mm}
\approx 46.3~\mathrm{mm},
\]
and hence
\[
D_{\mathrm{REV}}=2r_{\mathrm{REV}}\approx 92.7~\mathrm{mm}.
\]
Two points were emphasized.

First, the inferred $D_{\mathrm{REV}}$ was found to be a substantial fraction of the specimen
diameter,
\[
\frac{D_{\mathrm{REV}}}{D_{\mathrm{s}}}\approx \frac{92.7}{180}\approx 0.52,
\]
which was consistent with the qualitative challenge posed by \textit{Thalassinoides} fabrics: when a connected
burrow network exhibits centimeter-scale connectivity, a standard core diameter can sample only a limited
number of the longest-wavelength features, and representativeness in the transverse direction is then governed
by a small set of dominant structures.

Second, the radius estimate was interpreted as a statistically motivated \emph{minimum} diameter for stable
second-order descriptors of the detrended field. For diameters below $D_{\mathrm{REV}}$, the low-wavenumber
content was under-sampled, and the inferred correlation structure (and any downstream connectivity-driven property estimate) remained diameter-dependent. For diameters above $D_{\mathrm{REV}}$, progressively less new
low-$k$ information was added as the cylinder was enlarged, and the covariance/spectrum descriptors were thus
expected to stabilize within the prescribed tolerance.

\subsection{Combined REV estimate and practical implications}
Taken together, the axial window selected by the detrending procedure (Sec.~\ref{sec:detrend}) and the
transverse diameter selected by the $\widehat{C}$-stability criterion (Sec.~\ref{sec:Chat}) defined a practical
REV-sized extraction cylinder for the present sample, with $D_{\mathrm{REV}}\approx 93~\mathrm{mm}$ and
$H_{\mathrm{REV}}\approx 83~\mathrm{mm}$. 
These values were interpreted as the
smallest cylinder dimensions for which reproducible second-order structure
(covariance and low-$k$ spectral content) was obtained for the detrended
burrowsity field under further enlargement of the domain. In this sense,
$D_{\mathrm{REV}}$ provided a practical minimum field of view for downstream
digital-rock analyses controlled by long-wavelength connectivity, including
correlation-length reporting and the construction of simulation domains in
which a small number of connected burrow junctions could otherwise bias
effective-property estimates.

Rather than stating only that a representative subvolume was selected, the
workflow reports the analyzed field \(B(\mathbf{x})\), the segmentation rule,
the voxel spacing, the detrending window \(w^\star\), the low-wavenumber
criterion, the plateau/onset wavenumber \(k_0\), and the final physical
dimensions \(D_{\mathrm{REV}}\) and \(H_{\mathrm{REV}}\).
This information enables independent reproduction of the analysis scale and clarifies the distinction
between convergence of the mean fraction and convergence of the correlation structure.
Such reporting is particularly important in digital rock physics because two
subvolumes with similar porosity may have different long-wavelength
connectivity and therefore different permeability or elastic-response
predictions.

The identified REV dimensions can be used to mitigate finite-size effects in
petrophysical property estimation. For example, porosity and permeability calculations performed through digital-rock physics on
subvolumes smaller than \(D_{\mathrm{REV}}\) or \(H_{\mathrm{REV}}\) may
under-sample the long-wavelength burrow connectivity and therefore produce
biased estimates. The selected REV
provides a minimum admissible simulation support: property calculations should
preferentially be performed on domains at least as large as the reported REV,
or on several overlapping REV-sized windows when the full scan permits. If the
specimen contains only a small number of independent REV-sized domains, this
should be reflected in the uncertainty assigned to the resulting petrophysical
or geomechanical estimates.

Finally, the scope of the estimate was kept explicit. The reported values followed from (i) the detrending
window selection rule used to define $w^\star$ and (ii) the plateau identification used to determine $k_0$.
Both steps were transparent and reproducible (Figs.~\ref{fig:rolling window}--\ref{fig:C-hat}), but threshold
sensitivity remained. For practical reporting, it was therefore recommended that the selected $w^\star$ and the
identified $k_0$ be documented alongside the final physical REV dimensions, and that modest variations of these
intermediate quantities be treated as an uncertainty band on $D_{\mathrm{REV}}$ and $H_{\mathrm{REV}}$ rather
than as a contradiction of the method.

\subsection{Limitations of the low-wavenumber criterion}
\label{subsec:limitations}

The low-\(k\) stability criterion should be interpreted as a necessary
second-order representativeness test for the selected field, not as an
unconditional guarantee that all larger-scale heterogeneity has been sampled.
A plateau in \(\widehat{C}(k)\) may be reached even if heterogeneity exists at
scales larger than the available specimen, if the field is strongly anisotropic
but only isotropized descriptors are used, or if detrending removes a physical
large-scale component that is relevant for a particular property. Therefore,
the reported \(D_{\mathrm{REV}}\) and \(H_{\mathrm{REV}}\) are conditional on
the scanned sample, segmentation, detrending protocol, covariance estimator,
and tolerance. When possible, the selected REV should be checked against
independent or overlapping subwindows and against the convergence of downstream
properties such as permeability, elastic moduli, or connectivity metrics.

The workflow is not restricted to \textit{Thalassinoides} fabrics. For fractured
rocks, the analyzed scalar field could be a fracture/host indicator, aperture
map, fracture density field, damage variable, or segmented mineral/void phase.
For materials with a preferred fabric orientation, the isotropic covariance
\(C(r)\) and isotropic spectrum \(\widehat{C}(k)\) should be supplemented by
directional covariances \(C(\mathbf{r})\) or by spectra resolved along the
principal fabric directions. In that case, representativeness may be governed
by different REV lengths parallel and perpendicular to the dominant orientation.
For non-cylindrical or irregular samples, the same logic applies, but the
window function, mask correction, and admissible nested domains must be adapted
to the sample geometry. Thus, the transferable part of the method is the
sequence: define the relevant field, remove large-scale nonstationarity, compute
second-order descriptors on nested domains, and select the smallest domain for
which the low-wavenumber content is stable.

Sample geometry affects REV determination because it controls both the
available window functions and the longest wavelengths that can be sampled.
For the present CT core, nested inscribed cylinders are natural because they
respect the physical acquisition support and avoid including voxels outside
the specimen. A cubic sample would instead favor cubic, spherical, or
Cartesian-window analyses, with corresponding changes in the spectral window
and admissible Fourier modes. For irregular specimens, mask-aware covariance
estimators and boundary corrections would be required. Consequently, the
reported numerical value of \(D_{\mathrm{REV}}\) should not be interpreted as
a universal material constant independent of geometry; it is the minimum
representative diameter for the selected cylindrical support, field definition,
and preprocessing protocol.

\section{Conclusions}

In \textit{Thalassinoides}-bearing rocks, pore-scale variability is combined with a strongly connected,
centimeter-scale burrow architecture, so that the stability of image-derived descriptors can be
dominated by finite-size effects at wavelengths comparable to the scanned diameter. In the present
work, this issue was addressed by casting REV sizing as a \emph{second-order convergence} problem on
a finite CT cylinder. A reproducible burrowsity field was defined, and an explicit
detrending/denoising step was applied so that slow drift and structured CT artifacts were suppressed.
The resulting residual field was then used as input to a diameter-selection test in which
representativeness was enforced through the convergence of the low-wavenumber content of
\(\widehat{C}\) on nested cylinders.

When the protocol was applied to the present dataset, an REV-consistent axial extent of
approximately \(H_{\mathrm{REV}}\approx 83~\mathrm{mm}\) was obtained from the axial detrending of the
slice-wise phase-fraction signal, while an REV-scale diameter of \(D_{\mathrm{REV}}\approx 93~\mathrm{mm}\)
was indicated by the \(\widehat{C}\)-based stability criterion. When interpreted jointly, these values
were taken to define a recommended \emph{minimum} cylindrical analysis domain for correlation-descriptor
reporting and for downstream digital-rock workflows in which long-wavelength connectivity exerts
dominant control.
Beyond the present specimen, the main outcome was methodological: by linking REV sizing to the
convergence of covariance/spectral descriptors---rather than to a particular averaged observable---the
mechanism responsible for diameter dependence in connectivity-dominated media was targeted directly.
In future work, the diameters selected by \(\widehat{C}\) will be benchmarked against the convergence of
transport and mechanical observables (e.g., directional permeability estimates or elastic responses),
and sensitivity to segmentation choices and residual artifact levels will be quantified.

\section{Data availability}
MATLAB script used to generate the numerical results and figures in this work, together with example input and output data, are openly available in the GitHub repository
\url{https://github.com/quantumfi/REV_sizing}.

\begin{acknowledgments}
We thank Khalid Abdelbasit and Hassan Eltom for their help with fieldwork, and Hani Salman Al Mukainah at KFUPM for assisting with the CT-scan acquisition.
\end{acknowledgments}

\bibliography{main}

\end{document}